# A feasibility study of treatment verification using EPID cine images for hypofractionated lung radiotherapy[§]

**Xiaoli Tang**[1,2], **Tong Lin**[1,3], and **Steve Jiang**[1]

[1]Department of Radiation Oncology, University of California San Diego, La Jolla, CA 92093, USA
[2]Department of Radiation Oncology, University of North Carolina, Chapel Hill, NC 27514, USA
[3]Key Laboratory of Machine Perception (Ministry of Education), School of EECS, Peking University, Beijing 100871, China

E-mail: sbjiang@ucsd.edu

**Abstract.** We propose a novel approach for potential on-line treatment verification using *cine* EPID (Electronic Portal Imaging Device) images for hypofractionated lung radiotherapy based on a machine learning algorithm. Hypofractionated radiotherapy requires high precision. It is essential to effectively monitor the target to ensure that the tumor is within the beam aperture. We modeled the treatment verification problem as a two-class classification problem and applied an Artificial Neural Network (ANN) to classify the *cine* EPID images acquired during the treatment into corresponding classes—with the tumor inside or outside of the beam aperture. Training samples were generated for the ANN using digitally reconstructed radiographs (DRRs) with artificially added shifts in tumor location—to simulate cine EPID images with different tumor locations. Principal Component Analysis (PCA) was used to reduce the dimensionality of the training samples and *cine* EPID images acquired during the treatment. The proposed treatment verification algorithm was tested on five hypofractionated lung patients in a retrospective fashion. On average, our proposed algorithm achieved a 98.0% classification accuracy, a 97.6% recall rate, and a 99.7% precision rate.

## 1. Introduction

In the United States, lung cancer is the second most prevalent cancer and the leading cause of cancer death, accounting for about 30% of all cancer mortality (Jemal *et al* 2005). Hypofractionated lung radiotherapy is being increasingly employed as an alternate modality for the treatment of primary and secondary lung cancers. This therapy has the important advantages of allowing shortened treatment times while delivering higher effective radiobiological doses. However, normal tissues surrounding the tumors are also exposed to high-dose levels of radiation. Furthermore, cancerous tissue can occasionally move outside the irradiation field, e.g. when the patient has sudden irregular breathing or episodes of coughing. Under these circumstances, malignant tissue will be missed, and even more normal tissue than planned will be irradiated. A very large fractional dose (e.g., 10 Gy per fraction) is commonly applied in hypofractionated lung radiotherapy. This is in many ways an ablative therapy, both to the tumors and to the normal tissues surrounding them. Consequently, the precision requirement of hypofractionated lung radiotherapy is high. It is absolutely critical to effectively monitor the target to ensure maximal irradiation of the tumor with minimal irradiation of surrounding normal tissue.

---

[§] This work was first presented at the Seventh International Conference on Machine Learning and Applications, San Diego, CA USA, December 11-13, 2008.

The major uncertainty in treating lung cancer is the respiratory lung tumor motion, which can be clinically significant for some patients (e.g., of the order of 2 – 3cm) (Jiang 2006). This uncertainty must be dealt with when delivering hypofractionated lung radiotherapy. Typically, margins are added to accommodate respiratory motion. However, even with margins, tumors, or portions of them, will occasionally move outside the irradiation field. Abrupt coughing, dramatically changing breathing patterns, and sudden occurrences of pain, can all occur during treatment. Any one of these events can result in moving the tumor (or portions of it) outside the irradiation field. It is therefore, critically important to constantly monitor the patients' treatment—and when the tumor is detected outside the irradiation field, the treatment must be interrupted. The treatment should be resumed only when the tumor returns to the irradiation field or, in extreme cases, after patient re-setup.

EPID acquisition in *cine* mode does not require any additional radiation dose, and yet the technique generates images that carry valuable information indicating tumor position. Several methods for monitoring radiation therapy have been developed using *cine* EPID images, with or without implanted markers.

Berbeco *et al* developed a matching technique for respiratory-gated liver radiotherapy treatment verification with an EPID in *cine* mode (Berbeco *et al* 2005, Berbeco *et al* 2007). Implanted radio opaque fiducial markers inside or near the target were required for this technique. The markers were contoured on a planning CT set, enabling users to create digitally reconstructed radiographs (DRRs) for each treatment beam. During the treatment, a sequence of EPID images could be acquired without disrupting the treatment routine. Implanted markers were visualized in the images and their positions in the beam's eye view (BEV) were calculated off-line and compared to the reference position by matching the field apertures in corresponding EPID and DRR images. Tumor displacement was calculated for one patient with three implanted markers. The case study demonstrated the feasibility of the proposed method.

For lung cancer patients, implantation of fiducial markers is not widely acceptable due to the risk of pneumothorax (Laurent *et al* 2000, Arslan *et al* 2002, Geraghty *et al* 2003, Topal and Ediz 2003, Berbeco *et al* 2005). Arimura *et al* considered using *cine* EPID images for measurement of displacement vectors of tumor positions in lung radiotherapy without implanted markers (Arimura *et al* 2007). A template matching technique based on cross-correlation coefficients was proposed to calculate the similarity between a reference portal image and each *cine* EPID image. 5 patients with non-small cell lung cancer and one patient with metastasis were included for a validation study. The proposed method worked well for 4 cases but not well for the other 2.

To develop a more robust system, we propose an alternative approach for treatment verification of hypofractionated lung radiotherapy using *cine* EPID images without implanted markers. Artificial Neural Network (ANN) based technique will be developed to classify the *cine* EPID images into two classes: images with the tumor inside the radiation field and images with the tumor outside the radiation field.

This paper is organized as follows: section 2 will introduce methods and materials used in this work, including a brief introduction of ANN and a detailed description of how to apply ANN to our treatment verification problem. Section 3 presents experimental results. Section 4 will conclude this work and plan future work.

## 2. Methods and materials

The goal of on-line treatment verification is to monitor the tumor's position, to verify that it remains inside the radiation field (or beam aperture). If it is inside, the treatment can go on. Otherwise, the treatment beam should be turned off. This observation provides us with a clue that the on-line treatment verification problem can in fact be modeled as a classification problem. EPID *cine* images corresponding to the tumor inside the aperture can be treated as one class, and EPID *cine* images corresponding to the tumor outside the aperture can be treated as another. In this work, we apply a machine learning algorithm, ANN, for treatment verification. We will test its feasibility off-line retrospectively for hypofractionated lung radiotherapy.

*2.1. Artificial Neural Network*

An Artificial Neural Network is a mathematical model inspired by the way biological nervous systems process information. An ANN incorporates massively parallel systems with large numbers of interconnected simple processors, and it can solve many challenging computational problems. For a classification problem, an ANN will learn examples (training samples) of each class to extract corresponding patterns and detect unique trends. A trained ANN can therefore classify new samples into corresponding classes with high accuracy. More details on applying ANN on cancer research can be found in work by (Naquib and Sherbet 2001).

*2.2. Training ANN*

The first step of using any ANN is learning, training from samples. A trained neural network can be thought of as "expert" in the category of information it has been given to analyze. In our application of hypofractionated lung radiotherapy verification using *cine* EPID images, the ideal training samples would naturally be *cine* EPID images. There are, however, two problems. First, the ANN requires a large number of training samples to achieve reasonable results, and there aren't enough *cine* EPID images generated during treatment to meet that standard. Second, to be able to verify the treatment using *cine* EPID images during treatment, the ANN training has to be completed <u>before</u> the treatment, when the *cine* EPID images are not yet available. For these reasons, *cine* EPID images cannot be used as training samples.

We generate training images from DRR—to simulate *cine* EPID images with various artificially altered tumor locations. The DRRs were created in the BEV for each field. The field edges (MLC contours) were superimposed on these images. The first image in Figure 1 illustrates an example of DRR of a treatment field. The solid red contour is the MLC contour. By shifting the MLC contour, the sub-image defined by the contour changes accordingly. If each sub-image is treated as a simulated *cine* EPID image, we can simulate *cine* EPID images with different tumor locations. Once again, consider the first image of Figure 1: the blue and green dashed contours are two examples of MLC contour at different locations. Tumor locations are different in the sub-images outlined by the corresponding contours. In this fashion, we can simulate *cine* EPID images with different tumor locations. The two images on the right side of Figure 1 are the enlarged versions of the sub-images defined by the blue and green dashed contours in the first image, respectively. If we limit the MLC contour to move inside of an $m \times n$ pixel sized window at the step size of one pixel, we can generate $mn$ training images.

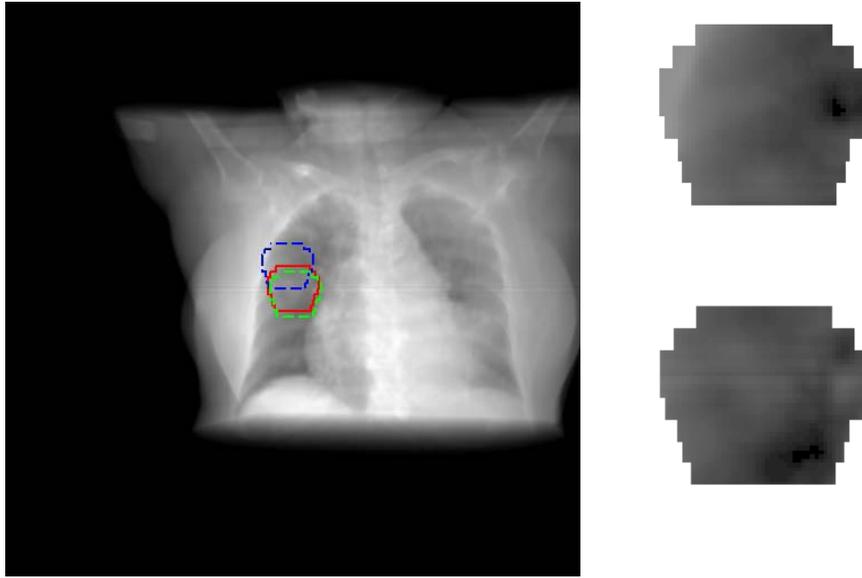

**Figure 1.** The image on the left: DRR of a treatment field. The solid red contour is the MLC edge. The blue and green dashed contours define two examples of the training images for ANN. The sub-images inside the blue and green contours have been enlarged and displayed on upper right and lower right, respectively. All images are original without any image processing.

The clinical target volume (CTV) is defined by the physician on 4DCT and is projected onto each DRR. Based on the location of the CTV, we can calculate what percentage of the tumor is in the beam aperture of each training image. With a user-defined threshold $p\%$, we associate class 1 to the training image if more than $p\%$ of the tumor is in MLC and class -1 otherwise.

From the total of $mn$ training images, the ANN can learn what features indicate class 1 and what features indicate class -1. It can create its own organization or representation of the information it receives during the learning time. The trained network can therefore analyze the *cine* EPID images obtained during the treatment and classify them into the corresponding class 1 or -1. It is worth to emphasize again that we used DRR to simulate *cine* EPID images for training, and the classification was done on real *cine* EPID images.

*2.3. Image processing*

As we stated above, we used DRR instead of *cine* EPID images for neural network training. However, DRR and *cine* EPID images are different image modalities. Their pixel resolutions might be different, and their intensity values might be in different ranges. To enhance the ANN's performance, we applied image processing techniques to the DRR to make DRR and *cine* EPID images more closely resemble each other. First we either sub-sample or interpolate DRR to make its resolution the same as *cine* EPID image, depending on the original resolution of DRR. Then, histogram equalization was applied on each DRR and *cine* EPID image to enhance image contrast. Finally, the intensity value of DRR was mapped to the same range of *cine* EPID images. Based on our experience, pre-processing the images can significantly improve the ANN performance.

*2.4. Principal Component Analysis (PCA)*

A typical *cine* EPID image might have an approximate size of $100 \times 100$ pixels. This means the dimensionality of a training sample would be $100 \times 100 = 10,000$. Significant computational time and resources would be needed to train the ANN with these high dimensional samples. This is simply not practical for on-line treatment verification. Consequently, PCA was applied to reduce the dimensionality of the training images. PCA is a classical statistical method. It involves a mathematical procedure that transforms original correlated variables into a small number of uncorrelated variables called principal components. The first principal component accounts for as much of the variability in the data as possible, and each succeeding component accounts for as much of the remaining variability as possible. In our application, we keep the first 15 principal components. Training 451 images with a reduced dimensionality of 15 using an un-optimized MatLab program on a regular PC (Intel Xeon CPU with 2 GB RAM) resulted in a running time of less than 3 seconds, demonstrating the effectiveness of this refinement.

**3. Experimental results**

All hypofractionated lung radiotherapy patients included in this study were treated on a Varian Trilogy Linac (Varian Medical Systems, Palo Alto, CA, USA) equipped with an electronic portal imaging device. During the treatment, the EPID was set to acquire images in the *cine* mode at a frame rate of 0.625 Hz. Figure 2 shows two sample *cine* EPID images. The corresponding DRR was shown in Figure 1.

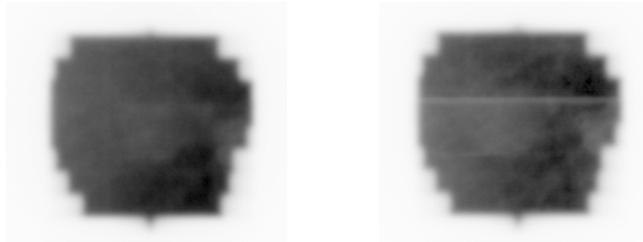

**Figure 2.** Examples of two original *cine* EPID images.

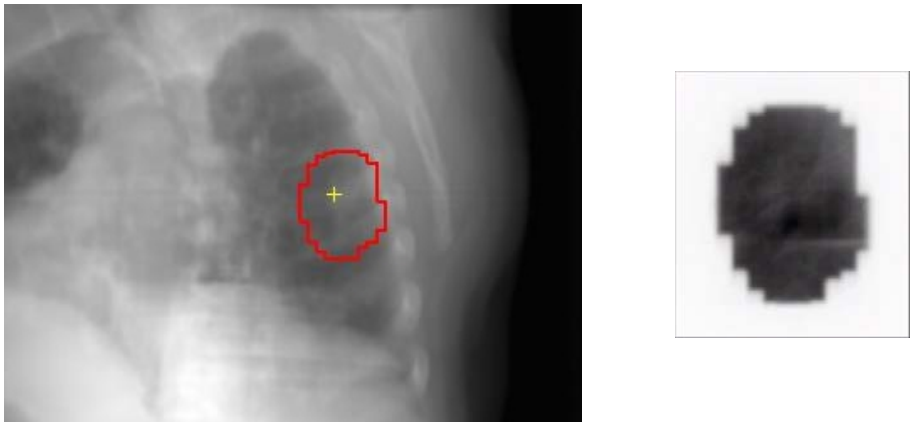

**Figure 3.** Original DRR and *cine* EPID image.

Five patients each treated with 4 or 5 fractions were included in our feasibility study. Table 1 lists the related patient information. The average patient age is 70, and the average tumor volume is 10.04 cm$^3$. The total number of *cine* EPID images of each patient varies from 84 to 329 depending on the treatment time. The examples of the DRR and *cine* EPID images of patient 1 were displayed in Figure 1 and Figure 2. Figure 3 shows examples of the DRR and *cine* EPID image of patient 4. The red contour on the DRR defines the bean aperture. Window size of $40 \times 10$ pixels was used on each DRR to generate the training images. A radiation oncologist read the *cine* EPID images and manually classified them into classes 1 and -1, and this serves as our ground truth. A three level feed-forward back propagation neural network was used. The network has a single hidden layer of 10 neurons, and the network is trained for up to 50 epochs to a error goal of 0.01. The ANN was applied on the training images to build the neural network. We set in this study—if more than 95% of the tumor projection is inside the irradiation field, the corresponding training image is said in class 1. For each treatment field, one neural network needs to be built.

**Table 1.** Patient information.

| Patient number | Age | Gender | Tumor location | Tumor volume (cm$^3$) |
|---|---|---|---|---|
| 1 | 77 | F | RUL | 5.02 |
| 2 | 75 | M | RUL | 17.23 |
| 3 | 65 | M | RUL | 2.23 |
| 4 | 46 | F | LLL | 1.39 |
| 5 | 85 | F | RLL | 24.35 |
| Average | 70 | - | - | 10.04 |

We would like to test if the proposed algorithm can successfully classify the *cine* EPID images corresponding to tumor outside the irradiation field obtained during the treatment. Those *cine* EPID images are category -1 images and also called true-negative images. However, since we were using real patient data retrospectively, we had very limited number of category -1 images. Therefore we simulated category -1 *cine* EPID images using the existing patient images.

Two patients whose tumor sizes were relatively small and treatment margins were large were selected for the simulation of category -1 images. For each treatment field, first we manually generated a sub-field inside the original field. As shown in figure 4 (a), the yellow contour defines the sub-field which covers the tumor. Then we trained the ANN model based on this new simulated sub-field MLC contour. Since the sub-field is much smaller than the original field, there was enough room to simulate other sub-fields with the same size but different locations. Figure 4 (b) and (c) are two examples of sub-fields with different locations, and they are examples of simulated category -1 *cine* EPID images. Figure 4 (b) corresponds to the situation that approximately half of the tumor is outside of the beam aperture, and figure 4 (c) corresponds to the situation that most of the tumor is not in the sub-field. The test *cine* EPID images include the simulated category -1 images as well as the category 1 images—images corresponding to tumor inside the treatment sub-field as shown in Figure 4 (a). The trained ANN model will classify the test images into corresponding categories. The two cases with simulated category -1 *cine* EPID images are called case 6 and 7.

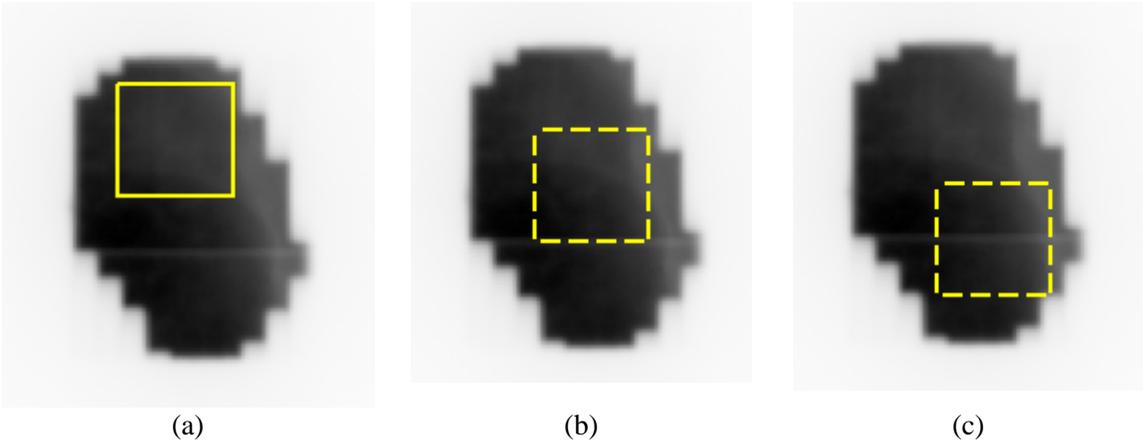

(a)                      (b)                      (c)

**Figure 4.** An example of generating sub-field. (a) The generated sub-field is defined by the yellow contour; (b) and (c) Two examples of simulated true negative *cine* EPID images.

We measure the accuracy, recall rate and precision of the classification results. They are defined as:

$$\text{accuracy} = \frac{\text{true positive} + \text{true negative}}{\text{all}} \quad (1)$$

$$\text{precision} = \frac{\text{true positive}}{\text{true positive} + \text{false positive}} \quad (2)$$

$$\text{recall} = \frac{\text{true positive}}{\text{true positive} + \text{false negative}} \quad (3)$$

Accuracy measures the degree of exactness or fidelity, precision measures the degree of reproducibility, and recall measures the degree of completeness. False positive is defined as when the tumor projection was not in the beam aperture, but the classification result was category 1. False negative is defined as when the tumor projection was in the beam aperture, but the classification result was category -1.

Table 2 lists the classification results of all five patients and case 6 and 7. Each number was averaged over all treatment fields. The average accuracy, precision, and recall numbers over all the patients are also listed in the last row of the table.

**Table 2.** Classification results in percentage.

| Patient | Accuracy | Precision | Recall |
|---|---|---|---|
| 1 | 94.12 | 100.00 | 94.12 |
| 2 | 99.87 | 98.86 | 98.89 |
| 3 | 97.00 | 100.00 | 97.00 |
| 4 | 98.86 | 99.87 | 98.89 |
| 5 | 98.91 | 100.00 | 98.91 |
| Case 6 | 99.00 | 100.00 | 98.77 |
| Case 7 | 98.13 | 99.08 | 96.43 |
| Average | **97.98** | **99.69** | **97.57** |

For all cases, the results are good; the numbers are in high nineties most of the time. Note all the precisions are either 100% or close to 100%. This means the reproducibility is high, and the proposed algorithm is stable. On average, the proposed algorithm achieved accuracy of 97.98%, precision of 99.69%, and recall of 97.57%.

## 4. Conclusion and future work

We have proposed a novel approach for on-line treatment verification for hypofractionated radiotherapy. The DRR was used to simulate *cine* EPID images for the ANN training. Image processing techniques were applied on the DRR to make the DRR and *cine* EPID images closely resemble each other. The PCA was also applied on training samples and *cine* EPID images acquired during the treatment to reduce their dimensionality in order to shorten the process time. We have tested our proposed algorithm on seven hypofractioned lung patient cases off-line in a retrospective fashion. The average accuracy and recall numbers are high, and the average reproducibility is close to 100%.

We intend to achieve even better results. More sophisticated image processing techniques will be applied to preprocess the DRR. We have already experienced a significant performance boost from pre-processing the images with the techniques described. Better image processing techniques should bring the classification accuracy rate even higher. All the ANN parameters were not optimized. We will investigate different combinations of parameters to find the set that yields the best performance. We used an empirical number 15 of principal components in this study. Research will be done to minimize the number of principal components while still achieve comparable results. Now we use kV beam DRR. We are developing software to generate MV beam DRR with scattering effect which will better resemble *cine* EPID images obtained during the treatment. The number of cases tested is relatively small. We are collecting more patient data, hopefully with implanted fiducial markers, to further validate our proposed algorithm. Patients with variety of tumor volumes and locations will be collected. The performance of the algorithm will be analyzed based on tumor volume and location. Comparison will be made on different tumor volumes and locations.

10

15